\documentclass[aps,prb,twocolumn,showpacs,showkeys,preprintnumbers,byrevtex]{revtex4}
\usepackage{amsmath,amssymb}
\usepackage{graphicx}
\makeatletter
\def\frontmatter@thefootnote{\@alph\c@footnote)}%
\makeatother

\begin{document}
\preprint{\texttt{mpi-pks/0203009}}

\title{A pure-carbon ring transistor: The r{\^o}le of topology and structure}
\author{Gianaurelio Cuniberti%
\footnote{E-Mail:~\texttt{cunibert@mpipks-dresden.mpg.de}}}
\author{Juyeon Yi}
\author{Markus Porto}
\affiliation{Max-Planck-Institut~f{\"u}r~Physik~komplexer~Systeme,
             N{\"o}thnitzer~Stra{\ss}e~38, 01187~Dresden, Germany}

\date{March 27, 2002}

\begin{abstract}
We report results on the rectification properties of a carbon nanotube (CNT)
ring transistor, contacted by CNT leads, whose novel features have been
recently communicated by Watanabe \textit{et al.} [Appl.\ Phys.\
Lett.~\textbf{78}, 2928 (2001)]. This paper contains results which are
validated by the experimental observations. Moreover, we report on additional
features of the transmission of this ring device which are associated with the
possibility of breaking the lead inversion symmetry. The linear conductance
displays a ``chessboard''-like behavior alternated with anomalous zero-lines
which should be directly observable in experiments. We are also able to
discriminate in our results structural properties (quasi-onedimensional
confinement) from pure topological effects (ring configuration), thus helping
to gain physical intuition on the rich ring phenomenology.
\end{abstract}


\maketitle

Carbon materials are at the base of the many molecular electronics achieved
goals.\cite{Service01} The discovery of the C$_{60}$ molecule\cite{KHOCS85}
and, later on, of carbon nanotubes (CNTs)\cite{Iijima91} provided
experimentalists with nanometer-scale materials with exceptional electronic and
mechanical properties.\cite{DDE96} The adoption of an individual C$_{60}$
molecule connected to gold electrodes made possible to measure transistor like
features.\cite{PPLAAMcE00} In a parallel development, also logic circuits with
field-effect transistors based on single CNTs have been
probed.\cite{BHND01,RKJTCL00} In most of the aforementioned experiments, the
focus has been centered on carbon-based molecules, C$_{60}$, single-wall (SW)
or multi-wall (MW) CNTs, bridged between bulky leads.

\begin{figure}[b]
\centerline{\includegraphics[scale=0.55]{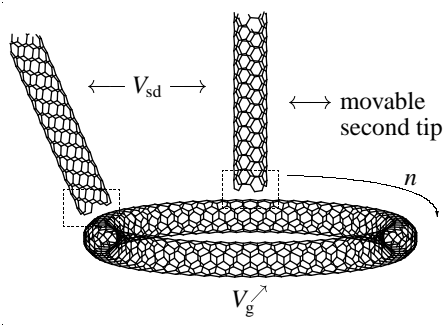}}
\caption{\label{fig:sketch}A pure-carbon ring transistor. The two semi-infinite
CNT leads can scan the upper surface of the CNT ring. The latter is supposed to
lie on a surface imposing a gate voltage $V_{\mathrm{g}}$, fixed with respect
to the second lead as in Ref.~\onlinecite{WMSS01}. The quality and nature of
the contacts (dashed boxes) is discussed in the text.}
\end{figure}

However, CNTs have been also shown useful as wiring elements,\cite{YWTA01}
e.g., when employed to enhance the resolution of scanning tunneling microscope
(STM) tips.\cite{HCL99,WJWCL98} Indeed, in a pioneering experiment published in
this journal, Watanabe \textit{et al.}\cite{WMSS01} managed to contact a CNT
ring to CNT-STM tips. Their apparatus, which is schematically drawn in
Fig.~\ref{fig:sketch}, realizes a pure-carbon transistor. The two CNT leads,
biased by a potential $V_{\mathrm{sd}}$, are the source and the drain for the
current flowing through the CNT-ring; the latter is sitting on a surface which
fixes the gate voltage to $V_{\mathrm{g}}$. This system, which is the object of
our theoretical investigation, exhibits an interesting variety of effects
mainly due to the ring topology and to the underlying carbon nanostructure. Our
results are validated by the experiment of Watanabe \textit{et
al.}\cite{WMSS01} but show interesting additional phenomena which should be
directly detectable in future experiments.

We treat the transport problem in the system at hand by calculating the
transmission $t$ through an armchair $(\ell, \ell)$ CNT ring within the
Landauer approach.\cite{Landauer57} The relation between the Hamiltonian of the
system and the transmission\cite{FL81} is given by means of the retarded Green
function $G^{\mathrm{ring}} = \left( E + \mathrm{i} 0^+ - H^{\mathrm{ring}}
\right)^{-1}$, where $H^{\mathrm{ring}}$ is the ring $\pi$ electron
Hamiltonian. Provided that the system has atomic contacts, one can show that
the calculus of the transmission simplifies to
\begin{equation} \label{eq:transm-mit-spectral-densities}
t_{n} (E) =4 \ \Delta_{11} (E) \Delta_{nn} (E) \left| G^{\mathrm{ring}}_{1n}
\left( E \right) / \mathrm{det} \left( {Q} \right) \right|^2,
\end{equation}
where $\Sigma = \Lambda - \mathrm{i} \Delta$ is the semi-infinite armchair
$(\ell,\ell)$ CNT lead self-energy, ${Q}= {I} - {\Sigma} \;
{G}^{\mathrm{ring}}$, and $I$ is the unity matrix. Recently, both $\Lambda$ and
$\Delta$ have been obtain analytically.\cite{CFR01c} We assume that the fixed
first lead has an atomic contact to the first ring atom of the upmost
circumference line spanned by the second lead. The position of the latter is
determined by the second contacted atom $n=2,\dots,2 c$ (see
Fig.~\ref{fig:sketch}). The total number of upmost atoms, $2c$, is two times
the number of unit cells, $c$, due to the choice of the armchair tubes. We can
span contacts (dashed area in Fig.\ref{fig:sketch}) ranging from a single
atomic contact to all surface atoms of the CNT leads contacted to the
lead.\cite{CFR01c} For sake of simplicity we present here only atomic contacts
results. The numerical computational task is thus shifted to the calculus of
the matrix element of $G^{\mathrm{ring}}$ between the two contacted atoms,
$G^{\mathrm{ring}}_{1n}$, and the determinant of the matrix $Q$; this problem
grows as $N^4$, $N=2 c \ell $ being the number of atoms in the CNT ring.

\begin{figure}[b]
\centerline{\includegraphics[scale=0.55]{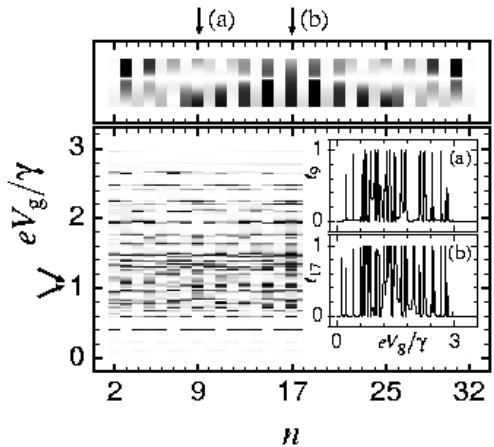}}
\caption{\label{fig:transm_CNT}Density plot of the linear transmission $t_n
(E\to 0)$ as a function of the position $n$ and of the gate voltage
$V_{\mathrm{g}}$. In this scale white corresponds to 0 and black to 1. In the
insets, the explicit gate voltage dependence of the transmission is shown for
(a) $n=9$ (90 deg) and for (b) $n=17$ (180 deg). The top panel illustrates a
blow up of the transmission in a small gate voltage window (indicated by tilted
arrows) where zeros in the transmission occur (a better resolution  figure is
available upon request).}
\end{figure}

Fig.~\ref{fig:transm_CNT} shows the dependence of the linear transmission on
the position of the second lead, $n$, and on the gate voltage,
$V_{\mathrm{g}}$. The latter is in units of the hopping parameter, $\gamma=2.6$
eV, between neighboring carbon atoms. The number of unit cells considered in
the ring is $c=16$ (i.e., $32$ upmost atoms) and the size of both the ring and
lead tubes are fixed by imposing $\ell=4$. The density plot appears quite
complex and reflects the presence of van Hove singularities in the system (see
the insets of Fig.~\ref{fig:transm_CNT}) but certain regularities emerge. At a
fixed gate voltage the position dependence of the transmission exhibit a
``chessboard''-like behavior. The latter is broken for particular values of the
gate voltage where zeros in the transmission occur. This is a typical
topological effect which has been shown analytically for one-dimensional
rings.\cite{YCP02} These zeros are a direct consequence of the breaking of the
inversion symmetry and are either due to destructive interference events or due
to the matching between the gate voltages and particular eigenvalues of the
system.\cite{YCP02}

With the knowledge of the transmission function $t_n (E)$, it is
straightforward to evaluate the $I$-$V$ characteristics by applying the
standard formalism based on the scattering theory of transport\cite{Datta99}
\begin{equation} \label{IV}
I_n = \frac{2e}{h}
 \int_{-\infty}^{\infty} {\rm d}E \; t_n (E) [f_{\rm L}(E)-f_{\rm R}(E)] .
\end{equation}
Here $f_{\rm L,R} (E)=\{{\rm exp}[(E-\mu_{\rm L,R} )/k_{\rm B}T] +1\}^{-1}$ is
the Fermi function, $\mu_{\rm L}$ and $\mu_{\rm R}$ are the electrochemical
potentials of the two metal electrodes, whose difference is fixed by the
applied bias voltage $V_{\mathrm{sd}}$. The room temperature ($k_{\rm B}T
\simeq 26~\mathrm{meV}$) will be considered in the numerical results, in order
to keep consistency with the experiment.\cite{WMSS01}

\begin{figure}[t]
\centerline{\includegraphics[scale=0.55]{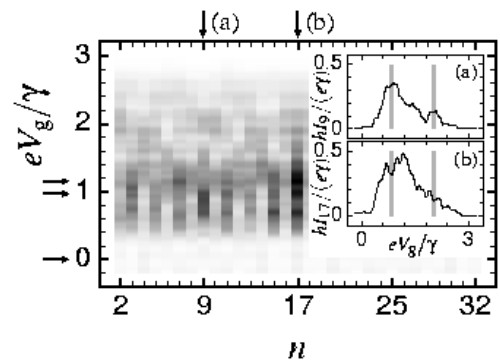}}
\caption{\label{fig:STM}Density plot of the current (STM images) at
$V_{\mathrm{sd}}=1~\mathrm{V}$ as a function of the position $n$ and of the
gate voltage $V_{\mathrm{g}}$. In this scale white corresponds to no current
and black to maximum current. The horizontal arrows refer to the three gate
voltages $V_{\mathrm{g}} = 0, 2.5, 3~\mathrm{V}$ considered in
Ref.~\onlinecite{WMSS01} (Fig.~3) for producing the STM images. The two insets
show the details of the gate voltage dependence for the two configuration (a)
$n=9$ ($90$ deg) and (b) $n=17$ ($180$ deg), respectively. The gray bars
indicate the regions where $I_9 \ge I_{17}$ (a better resolution  figure is
available upon request).}
\end{figure}

\begin{figure}[t]
\centerline{\includegraphics[scale=0.55]{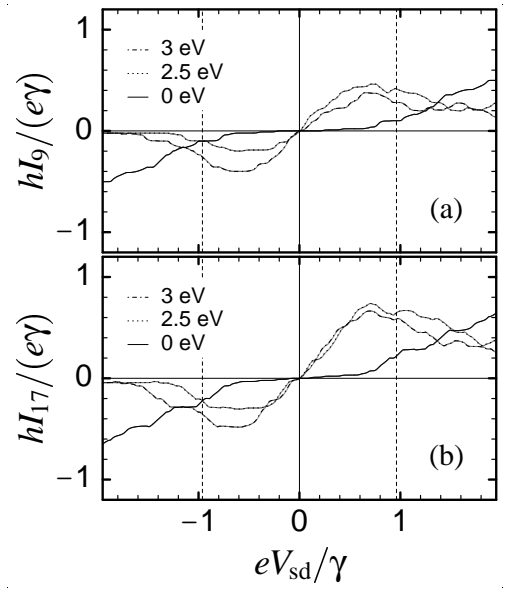}}
\caption{\label{fig:IV_CNT_180_90}Lead-orientation dependent rectification, for
three different gate voltages $V_{\mathrm{g}} = 0, 2.5, 3~\mathrm{V}$,
illustrated (a) for the $90$ deg configuration and (b) for the inversion
symmetric one.}
\end{figure}

Fig.~\ref{fig:STM} displays the current at $V_{\mathrm{sd}} = 1~\mathrm{V}$ for
different positions of the second CNT tip and for different gate voltages. This
quantity has been used to produce the STM images in Ref.~\onlinecite{WMSS01},
where three different values of gate voltages have been considered (here
indicated as the three horizontal arrows in Fig.~\ref{fig:STM}). The inversion
symmetric configuration shows a marked current density for wide ranges of gate
voltages, apart from two small intervals around 2.7~V and 5.8~V, respectively.
At these voltages, transport is much more favored in the lobes of the ring
which, for the choice of our size, emerge at the 90 deg configuration (see the
gray regions in the inset of Fig.~\ref{fig:STM}).

Fig.~\ref{fig:IV_CNT_180_90}~(b) shows the $I$-$V$ characteristics for the
inversion symmetry position of the second lead (180 deg). The rectification
effect given by the action of the gate potential are due to the finiteness of
the ring band which, under a gate shift, is pushed in a region without states.
Different is the situation in the 90 deg configuration
(Fig.~\ref{fig:IV_CNT_180_90}~(a)), where no marked asymmetry in the gated
$I$-$V$ curves can be detected. The vertical dashed lines in
Fig.~\ref{fig:IV_CNT_180_90} indicate the range of bias voltages investigated
in Ref.~\onlinecite{WMSS01}. Our calculations outside this region show dominant
negative differential conductance effects given by the action of the gate
voltage. The latter is responsible for shifting the CNT ring band in a energy
region where much fewer states are available for transport.

More detailed experimental studies on the evolution of the differential
conductance with a broken inversion symmetry might probe the striking scenario
of the dependence of the STM images on both position and gate voltage.
Moreover, this pure-carbon system, where electron transfer between leads and
molecule is approximately negligible, may turn out as the best playground to
study the effect of contacts on the overall transport measurement.

\begin{acknowledgments}
GC research at MPI is sponsored by the Schloe{\ss}mann Foundation.
\end{acknowledgments}

\end{document}